\documentstyle[12pt,aps]{revtex}
\newcommand{\be}{\begin{equation}}
\newcommand{\ee}{\end{equation}}
\newcommand{\ben}{\begin{equation*}}
\newcommand{\een}{\end{equation*}}
\newcommand{\bea}{\begin{eqnarray}}
\newcommand{\eea}{\end{eqnarray}}
\newcommand{\bean}{\begin{eqnarray*}}
\newcommand{\eean}{\end{eqnarray*}}

\begin{document}
\title{Dissipative Dynamics of an Open Bose Einstein Condensate}

\author{
F.T. Arecchi$^{\diamondsuit,\spadesuit}$, 
J. Bragard$^{\diamondsuit,\clubsuit}$
and L.M. Castellano$^{\diamondsuit,\heartsuit}$} 
\address{
$\diamondsuit$ 
Istituto Nazionale di Ottica, Largo E. Fermi, 6, I50125, Florence, Italy \\
$\spadesuit$
 also at Dept. of Physics, University of Florence, Florence,
Italy \\
$\clubsuit$ also at Dept. of Physics, University of Liege,
 Liege, Belgium \\
 $\heartsuit$ on leave from Dept. of Physics, University of Antioquia, 
 Medellin, Colombia \\
 email: arecchi@ino.it
 }

\date{\today}
\maketitle

\begin{abstract}

As an atomic Bose Einstein condensate (BEC) is coupled to a source of 
uncondensed atoms at the same temperature and to a sink  
(extraction towards an atom laser) the idealized description in terms of a 
Gross-Pitaevsky equation (GP) no longer holds. Under suitable physical 
assumptions we show that the dissipative BEC obeys a Complex Ginzburg 
Landau equation (CGL) and for some parameter range it undergoes a space 
time patterning. As a consequence, the density of BEC atoms within the 
trap displays non trivial space time correlations, which can be detected 
by monitoring the density profile of the outgoing atom laser.
The patterning condition requires a negative scattering length, as e.g. in
$^7$Li. In such a case we expect a many domain collapsed regime, rather than
a single one as reported for a closed BEC.

\end{abstract}

\section{ Introduction}

The BEC dynamics in an atomic trap is ruled by a GP 
\cite{GP1,DAL1} which in fact is a nonlinear Schr\"odinger equation (NLS)
describing a conservative motion.
Experimental evidence of BEC in a trap \cite{AND1,BRAD1,DAV1}
 confirmed qualitatively
a dynamical picture based on a GP description.
On the other hand, extraction of BEC-atoms toward an atom laser 
\cite{MEWES1,BLO1}
introduces a dissipation which must be compensated for by a transfer
from the uncondensed fraction of trapped atoms.
Those ones on their turn must be refilled by a pumping process which, in
the actual laboratory set ups is a discontinuous 
process \cite{MEWES1,BLO1}
but that we here consider as a continuous refilling, even though
no working scheme is available yet.
In Sec.\ref{sec2} we describe the addition of dissipative interactions through
coupled rate equations, as done by Kneer et al. \cite{KNEER1}.
In Sec.\ref{sec3} we provide the physical grounds for an additional space
dependent (diffusive) process and introduce an adiabatic 
elimination procedure,
whereby we  arrive at a closed equation which in fact is a CGL.
In Sec.\ref{sec4} we rescale the CGL around threshold for both 
positive ($^{87}$Rb) and negative ($^7$Li) scattering lengths, showing that
in the first case the BEC is stable against space time variations,
whereas in the second case the system can easily cross the instability barrier
(so-called Benjamin-Feir line \cite{BEN1,SHRA1,CHATE1}).
In Sec.\ref{sec5} we present numerical results showing that
in the unstable case, rather than collapsing into one singular spot
as in the isolated BEC, the open system presents many uncorrelated
domains (space-time chaos).
In Sec.\ref{sec6} we compare the strength of the 
nonlinear dissipative term introduced by us
with the 3-body recombination rate.


\section{The Dynamics of an Open BEC}
\label{sec2}

 We know\cite{GP1,DAL1} that a BEC is modeled  by the GP  
\be i\hbar
\frac {\partial \phi}{\partial t}= -\frac{\hbar^2}{2m}\nabla^2\phi 
+ V_{ext}\phi + g\mid\phi\mid^2\phi. 
\label{eq1}
\ee

where $\phi =\sqrt{\rho} e^{i\theta} $ is the macroscopic wave function 
describing the probability amplitude of the condensate,
$V_{ext}$ is the trap potential, shaped as a harmonic
oscillator with frequency $\omega$, and $g$ is the coupling constant 
for the nonlinear  (density $\mid\phi\mid^2$ dependent) self interaction.
 $g$ is proportional to the s-wave scattering length $ a_s $ 
 
\be 
g = \frac{4\pi\hbar^2}{m}a_s. 
\label{eq2}
\ee

We discuss   specific experimental situations concerning
$^{87}$Rb atoms ($m = 1.4 \, 10^{-25}$ kg, $ a_s = 5.77$nm) 
and $^7$Li atoms ($m = 0.115 \, 10^{-25}$ kg, $ a_s = -1.45$nm)
\cite{DAL1}.
For an anisotropic trap the frequency is  
$\omega = (\omega_x \omega_y \omega_z)^{\frac{1}{3}}$ 
 \cite{DAL1,AND1,BRAD1,DAV1}.
The above equation is formally a conservative non linear
Schr\"odinger  equation (NLS). Thus, it is a straightforward task to attach
 to a BEC all those space-time features familiar of a NLS as e.g. 
 solitary structures and vortices \cite{CIRAC1,KI1}, 
 which have been explored in the recent past for a NLS,
mainly in connection with pulse propagation in optical fibers 
\cite{HAS1,KUM1}.

On the other hand the idealized picture of a BEC in an
isolated system is in contrast with two physical facts, 
namely, \newline
i) The BEC is made
of that fraction of atoms which have collapsed into the ground state
($n=0$) of the harmonic oscillator trap potential;
these atoms interact via collisions with those ones which are distributed 
over the excited states $(n>0)$ of the trap. The uncondensed atomic density 
$n_u$ evolves in time  not only because of the coupling
with the condensed phase described by $\phi$, but also because trapping 
and cooling processes imply a feeding (pumping) at a local rate 
$ R (\vec{r})$,   the space
dependence accounts for the non uniformities of the pumping process
 as well as for
losses due to escape from the trap, at a rate $\gamma_u$. \newline
ii) In order to have an atom laser, a radio frequency (rf) field is applied 
to the trap. The rf changes the magnetic quantum number of
 the atoms' ground state, thus transforming the
trapping potential into a repulsive one and letting atoms escape  
from the BEC at  a rate $\gamma_c $ \cite{MEWES1,BLO1}.

Both i) and ii) have been 
modeled by Kneer et al \cite{KNEER1} by adding  dissipative terms to 
Eq.(\ref{eq1}) and coupling the
resulting equation with a rate equation for $n_u$. 
In a slightly different formulation,
 this   amounts to the
following  equations 

\be 
i\hbar\frac {\partial \phi}{\partial t}= 
-\frac{\hbar^2}{2m}\nabla^2\phi + V_{ext}\phi +
g\mid\phi\mid^2\phi -\frac{i}{2}\hbar\gamma_c\phi 
+ \frac{i}{2}\hbar{\Gamma} n_u \phi
\label{eq3} 
\ee 
and
\be \dot{n_u}= R(\vec{r}) - \gamma_u n_u - \Gamma n_u n_c. 
\label{eq4}
\ee 
here ${\Gamma }$ is the rate constant coupling the condensed field 
$ \phi $ with the uncondensed density $ n_u $, and $ n_c = \mid\phi\mid^2$ 
is the local density of the condensed phase.
We have modified the model of Ref.\cite{KNEER1} as follows.
 At  variance with \cite{KNEER1}, where Eqs.(\ref{eq3},\ref{eq4}) 
 were written for the overall atomic
 population $N_u$ over the whole  trap volume V, that is, 
 \be
 N_u = \int n_u d^3r,
 \label{eq5}
 \ee
 which is then coupled to
 \be
 N_c = \int n_cd^3r = \int\mid\phi\mid^2(r)d^3r,
 \label{eq6}
 \ee
here we prefer to deal with a local coupling.
In fact Eqs.(\ref{eq3},\ref{eq4}) as written above are more convenient, 
as they refer to a local interaction.
 The coupling rate of uncondensed to condensed atoms,
$\Gamma$ ($m^3 \, s^{-1}$), is given by the global rate used 
in \cite{KNEER1} which we call $\Gamma^{\prime}$, dividing
by the trap volume $V$
\be 
\Gamma^{\prime} = \frac{{\Gamma}}{V} \, (s^{-1}).
\label{eq7}
\ee
Furthermore our local feeding rate $R(r)$ is related to the overall 
rate $R_u$ of \cite{KNEER1} by 
\be 
R_u=\int R(r)d^3r.
\label{eq8}
\ee

 
\section{CGL Picture of the Open BEC} 
\label{sec3}
  
Eqs.(\ref{eq3},\ref{eq4})  were the basis of the model reported in 
Ref.\cite{KNEER1}. We wish to
improve that picture, based on the following considerations.
The uncondensed phase, $n_u$, is fed by a pumping process $R(r)$ 
which is in general non uniform, and
is locally depleted by its coupling with the condensed phase.
As a result, $n_u$ has a sensible
space dependence and hence it undergoes diffusion processes.
Precisely, by the fluctuation-dissipation theorem
\cite{KUB1,CAL1}, the diffusion in velocity is given by 
\be 
D_v = \gamma_u \frac{k_BT}{m} \, (m^2s^{-3}). 
\label{eq9}
\ee

The corresponding diffusion constant in real space will be 
\be
D_r = \frac{D_v}{\gamma_u^2}= \frac{1}{\gamma_u}\frac{k_BT}{m }
\, (m^2 s^{-1} ).
\label{eq10}
\ee

For $ ^{87}$Rb  at $T = 100$ nK , and for $\gamma_u = 5 \, 10^2 s^{-1}$
 (of the order of the
average trap frequency $\tilde{\omega} \approx  \pi  132 \, $Hz) 
\cite{KNEER1} this yields 
\be
D_r = 2 \, 10^{-8} \, (m^2s^{-1}).
\label{eq11}
\ee

Thus we must add the term $D_r \nabla^2 n_u$ to Eq.(\ref{eq4}).
Once the BEC has been formed, the escape rate $\gamma_c$ in 
Eq.(\ref{eq3}) is compensated
for by the feeding rate ${\Gamma} \, n_u$.  As we set  the BEC close
 to threshold,
because of critical slowing down, the $\phi$ dynamics will be much slower 
than the $n_u$ dynamics, thus we can apply an adiabatic elimination 
procedure \cite{HAK1}, find a quasi stationary solution for $n_u$ in 
terms of $\phi$ and replace it into Eq.(\ref{eq3}) which then
becomes a closed equation for $\phi$. 

We specify the above procedure by  the following steps. 
First, rewrite Eq.(\ref{eq4})
including diffusion  
\be 
\dot{n_u} = R(r)-\gamma_u n_u + D_r\nabla^2n_u - \Gamma
n_u\mid\phi\mid^2 .
\label{eq12}
\ee 
Next, we take its space Fourier transform. The linear terms are
trivial, whereas the nonlinear term should provide a convolution integral.
However, since the condensate has collapsed on a single mode $k=0$, 
then its Fourier transform is 
${\cal F}{\mid\phi(r)\mid^2}=\mid\phi\mid_0^2 \, \delta(k)$ 
and Eq.(\ref{eq12}) transforms as 
\be 
\dot{n_k}= R_k-(\gamma_u + D_rk^2)n_k - \Gamma n_k\mid\phi\mid_0^2. 
\label{eq13}
\ee 
where
$n_k = {\cal F}{n}_u $, $R_k = {\cal F}{R(r)}$. 
The adiabatic elimination procedure 
consists in taking the stationary solution of Eq.(\ref{eq13}) 
and replacing into Eq.(\ref{eq3}).
 The stationary solution of Eq.(\ref{eq13}) is 
 \be 
 n_k = \frac{\frac{R_k}{\gamma_u}}
 {1+ \frac{D_r}{\gamma_u}k^2 + \frac{\Gamma}{\gamma_u}\mid\phi\mid^2}. 
 \label{eq14}
 \ee 
 For long wavelength perturbations and far from saturation,
  the two additional terms in the denominator are less than unity.
   Here we consider a cylindrical volume with 
   $L_x = L_y$ = $5\mu$m, $L_z = 10\mu$m (V = 250($\mu$m)$^3$)
     containing a condensate of $N_c= 5 \, 10^4$ atoms at a 
     temperature $T =100n$K \cite{LENS}.
It follows that $k \approx O(1/L)\approx 0.5 \, 10^5m^{-1}$, 
and hence  $D_r k^2/\gamma_u< 1 $.
 We can then expand Eq.(\ref{eq14}) as 
 \be 
 n_k \approx \frac{R_k}{\gamma_u}(1-\frac{D_r}
 {\gamma_u}k^2 -\frac{\Gamma}{\gamma_u}\mid\phi\mid_0^2). 
 \label{eq15}
 \ee 
The inverse Fourier transform of Eq.(\ref{eq15}) is  an operator relation
 as  
 \be
   n_u = \frac{R}{\gamma_u} (1 + \frac{D_r}
   {\gamma_u}\nabla^2 - \frac{\Gamma}{\gamma_u}\mid\phi\mid^2). 
  \label{eq16}
  \ee 
  As we replace this expression into Eq.(\ref{eq3}),
  the operator $\nabla^2$ acts on its right upon
the space function $\phi$. 
By doing this, we  derive at a closed equation for $\phi$
which reads as  
\be
 \frac {\partial \phi}{\partial t}= -\frac{i}{\hbar}[GP] +
\frac{1}{2}(\frac{R}{\gamma_u}- \gamma_c)\phi 
+ \frac{RD_r}{2\gamma_u^2}\nabla^2\phi
-\frac{R \Gamma^2}{2\gamma_u^2}\mid\phi\mid^2\phi, 
\label{eq17}
\ee 
where the square brackets contain the right hand side of Eq.(\ref{eq1}). 
We now write explicitly the GP terms thus
arriving at the following CGL 
\be 
\dot{\phi} = [\frac{1}{2}(\frac{R\Gamma}{\gamma_u}
- \gamma_c) - \frac{iV_{ext}}{\hbar}]\phi 
+ (\frac{RD_r\Gamma}{2\gamma_u^2} 
+\frac{i\hbar}{2m})\nabla^2\phi - (\frac{R{\Gamma}^2}{2\gamma_u^2} 
+ i\frac{g}{\hbar})\mid\phi\mid^2\phi. 
\label{eq18}
\ee

The dissipative terms of Eq.(\ref{eq18}) represent
respectively: i) difference between gain and losses,
which implies a threshold condition; ii) a real
diffusion which implies a spread of any local perturbation;
iii) a real saturation term which provides a density dependent
gain saturation.


\section{Rescaled CGL near Threshold} 
\label{sec4}

We herewith list the numerical values as taken from the experiment
\cite{LENS} or from Ref.\cite{KNEER1}.
We refer to a trap volume
 $V=0.25 \, 10^{-15} $m$^3$; with a loss rate
 $\gamma_u = 500 \, s^{-1}$.
If we take the value \cite{KNEER1} 
$\Gamma^{\prime} \approx 7 \, s^{-1}$, then
$\Gamma = \Gamma^{\prime} V \approx 2 \, 10^{-15} \, m^3s^{-1}$.
Furthermore, a reasonable estimate for the
BEC escape rate toward the atom laser is
\cite{KNEER1} 
$\gamma_c \approx 50 \, s^{-1}$.
Therefore the threshold condition (gain=losses) 
 is fulfilled for
 \be
 R\Gamma/\gamma_u = \gamma_c 
 \label{eq19}
 \ee
 where $R \approx 12 \, 10^{18} \, m^{-3}s^{-1}$, corresponding to
$R_u = R V$ $\approx 3000 \, s^{-1}$.
Finally   we have for $^{87}$Rb \cite{DAL1}
$g/\hbar = (4 \pi \hbar a_s) /m =
0.48 \, 10^{-16}m^3s^{-1}$.

 We can  now write  the parametrized CGL equation (\ref{eq18}) in the
 dimensionless form
 \be
 \dot{\phi} =  \epsilon \phi + (1+i c_1)\nabla^2 \phi - (1 +
 ic_2)\mid\phi\mid^2\phi.
 \label{eq20}
 \ee
where we have used the dimensionless time 
\be 
\tau = \gamma_c t ,
 \label{eq21}
 \ee 
and the dimensionless space coordinates 
\be 
\frac{x}{l_{o}}, \, 
\frac{y}{l_{o}}, \, 
\frac{z}{l_{o}},
\label{eq22}
\ee 
where  
\be
 l_{o} = (\frac{R D_r \Gamma}{2 \gamma_u^2 \gamma_c})^{1/2} 
 \,  \approx 3 \, 10^{-6}m
 \label{eq23}
  \ee 
  is the characteristic length associated to the CGL dissipative
  dynamics, 
and the dimensionless condensated wave-function 
\be
|\tilde{\phi}|^2 = 
\frac{R \Gamma^2}{2 \gamma_u^2 \gamma_c} |\phi|^2 .
\label{eq24}
\ee
Note that tilde has been dropped in Eq.(\ref{eq20}).
It follows from Eq.(\ref{eq18}) that  
\be
\epsilon = \frac{1}{2}(\frac{R\Gamma}{\gamma_u\gamma_c}-1),
\label{eq25}
 \ee 
 \be
 c_1= \frac{\hbar \gamma_u^2}{m R D_r \Gamma},
 \label{eq26}
 \ee
 \be
 c_2= \frac{2 g  \gamma_u^2}{\hbar  R  \Gamma^2},
 \label{eq27}
 \ee
 are the significant parameters of Eq.(\ref{eq20}). They
 are  pure numbers.
 The term $-i V_{ext}/\hbar \, \phi =-i c_0 \, \phi $ in
  Eq.(\ref{eq18}) can be eliminated by
 a rotation transformation $\phi \rightarrow \phi \, e^{-i c_0 t}$.

We notice that Eq.(\ref{eq20}), derived by sound
physical assumptions, is far from being a purely conservative (GP)
 or purely dissipative (real Ginzburg Landau) equation,
 but it displays both characters. 

However the Benjamin-Feir instability condition \cite{BEN1}
\be
c_1 \, (-c_2) > 1
\label{eq28}
\ee
is not met by $^{87}$Rb ($c_1$=0.52, $c_2$=2.05) and its
dissipative CGL is fully inside the stable region. Hence the 
addition of dissipative terms may add interesting transient effects
but does not lead to substantial qualitative changes with respect
to the GP equation.
Quite different is the case of $^7$Li ($c_1$=0.04, $c_2$=-0.51).
 Indeed even though the values of $c_1$ and $c_2$ 
 just listed give a stable dynamics,
the fact that the scattering length is negative may lead to an instability
if the parameters of the open BEC are slightly changed, e.g. if 
$\gamma_c$ is reduced by a factor $10$
(which physically corresponds to a Li atom-laser
with weaker losses).
In such a case, we get 
$c_1$=0.4, $c_2$=-5.1 and the open BEC is in the unstable region;
we will denote this experimental situation by referring to
an open Li$^*$ BEC.

\section{Numerical simulations}
\label{sec5}

As we have shown in Sec.\ref{sec4}, the coefficients of the CGL
depend on the nature of the atoms forming the open BEC and also
depend on the characteristic working parameters of the open BEC.
Let us discuss the space time dynamics of the density of the condensed
phase $|\phi|^2$. To do this, we proceed to the numerical integration of 
Eq.(\ref{eq20}). The integration is performed on a two dimensional domain.
This corresponds to a cross section of the 3-D cigar shape where the 
condensation takes place. This is justified by the fact that 
$\omega_z >> \omega_x \, , \omega_y$.
The simulations are done on a 200x200 (Rb) or 256x256 (Li$^*$) grid starting
with an initial Gaussian distribution at the center of the domain.
The numerical integration code is based on a semi-implicit scheme in time with
finite difference in space. The chosen boundary conditions
(at $x= \pm L_x/2$ and $y= \pm L_y/2$) are 
\be
\frac{\partial \phi}{\partial n} = -0.1 \phi
\label{eq28b}
\ee
where $n$ is the normal at the boundary. Eq.(\ref{eq28b}) expresses the
condition of an isotropic output flux of the condensed BEC 
(in the ideal situation of zero-gravity).
The numerical coefficient on the right hand side
of Eq.(\ref{eq28b}) is the dimensionless ratio between
$\gamma_c$ and the velocity modulus of the condensed atoms,
easily evaluated from the ground state solutions of the 
harmonic oscillator \cite{DAL1}.
In Fig.1   four snapshots of $|\phi (x,y,t)|^2$ are shown
at different times for
the Rb case ($c_1=0.52$, $c_2=2.05$ and $\varepsilon=0.5$). The initial 
distribution evolves towards a stable 
quasi-uniform state. Fig.2 displays three cross section of Fig.1
at different times, the solid line corresponds to the final stationary state
and we observe the  nearly uniform condensate on the overall domain.
Fig.3 illustrates a quite different situation: The values are now
$c_1=0.4$, $c_2=-5.1$ and $\varepsilon=0.5$ which corresponds to
a Li$^*$ open BEC in the unstable region of use. 
The space-time chaotic dynamics
emerges after a short transient ($t<10$). 
Fig.4 confirms that $|\phi|^2$
 is no longer symmetric with respect to $x=0$ (the same holds for the
 y-axis).  Figure 5 is aimed to show the spatial 
 decorrelation of the signal
 when the condensate has entered the chaotic regime.

 The 1D spatial power spectrum of the $|\phi(x)|^2$ function is shown in
 the lower curve of Fig.5.,
  the upper curve (solid line) is 
 calculated by averaging the power spectrum of the function $|\phi(x)|^2$ 
 over a time interval from $t=500$ until $t=1000$ 
 (taking a sampling time  $\delta t= 10$)
 within which the dynamics is statistically stationary. 
 The results clearly indicates the large spatial 
 decorrelation of the signal inside the chaotic regime.
 Indeed, it is well known that the Fourier transform of a Gaussian
 function 
 $G \propto e^{-x^2/\sigma_x^2}$ is again a Gaussian function
  $\tilde{G} \propto e^{-k^2/\sigma_k^2}$ with $\sigma_k \sigma_x=1$.  
On Fig.5 it appears that the bandwidth in the Fourier space is
much larger in the chaotic regime 
than for the initial distribution, which means a decorrelation of 
$|\phi(x)|^2$ once the system becomes chaotic.

To give a quantitative feeling, in the case of 
Li$^*$, we have reduced $\gamma_c$ by a factor 10, which means
that the normalization length is increased by $\sqrt{10}$ 
with respect to Eq.(\ref{eq23}),
and it is $l_o^* \approx 10 \mu$m.
In the numerical calculations we have considered a trap
of linear size $L= 50 l_o^*$. Since the
ratio of the spectral widths between the
chaotic and the initial spectra is about 4 (estimated from
Fig.5), it results that the coherence length $\sigma_c$  in  
space time chaos is $\sim 1/4$ the length of the initial
Gaussian packet $\sigma_0$.
As seen from Fig.4 $\sigma_0 \cong 5 l_o^* \cong 1/10 L$,
 hence $\sigma_c \cong 1.2 l_o^* \cong 1/40 L$.
These numerical estimates agree with a densitometric analysis 
of Fig.3.

\section{Discussion and Conclusions}
\label{sec6}


Kagan et al. \cite{KAG1} have discussed the collapse of a BEC in
$^7$Li for a number of condensed atoms $N_0$ larger than the 
critical value $N_{crit}$, that is,
\be
N_{crit} \approx \frac{a_{ho}}{|a_s|}.
\label{eq29}
\ee
This relation for $N_{crit}$ is obtained by equating the level spacing
$\hbar \omega$ in the trap to the interparticle interaction energy
$n_0 |g| = N/V \, |g|$ where $g=4 \pi \hbar^2 a_s / m$ and 
$V \approx a_{ho}^3$.
Ref.\cite{KAG1} stabilizes the  GP via a dissipative term
corresponding to 3-body recombination processes.
This amounts to a correction corresponding to a 6th power term in
$\phi$ in a free energy potential.
The dissipative equation of Ref.\cite{KAG1} is then
\be
\dot{\phi} = -i [ GP ] - \xi |\phi|^4 \phi
+ \hbox{(pumping from the uncondensed portion)}.
\label{eq30}
\ee
In Eq.(\ref{eq20}), we have already treated the last term, here 
expressed in words, by the Kneer et al. approach \cite{KNEER1}.
Let us now compare the 5th power real damping entering  Eq.(\ref{eq30})
with the 3th power real damping of  Eq.(\ref{eq20}).
The cubic  term is of the form $G_3 \phi$,
where
\be
G_3= \frac{R \Gamma^2}{2 \gamma_u^2} |\phi|^2 
= \frac{\gamma_c \Gamma}{\gamma_u} |\phi|^2. 
\label{eq31}
\ee
Using the numerical values corresponding to the $^{87}$Rb
 we obtain $G_3 \approx 48 \, s^{-1}$ , for
 $^7$Li we have $G_3 \approx 1920 \, s^{-1}$.
The cubic rate ($G_3$) is a combination of the three characteristic rates
of an open BEC.
In a similar way we can introduce the rate
$G_5 = \xi |\phi|^4 $. Taking the numerical values
provided in Ref.\cite{KAG1} we have the following ratios
between the  two dissipation rates
\be
\frac{G_3}{G_5} = 
\left\{
\begin{array}{ll}
480 & \mbox{for } ^{87}\mbox{Rb} \\
2 \, 10^4 & \mbox{for } ^7\mbox{Li}
\end{array}
\right.
\label{eq32}
\ee
This result clearly indicates that for an open BEC, the
3-body recombination is negligible with respect to the saturation cubic term
that comes
from the coupling between the condensed and uncondensed phase of the
open BEC.

To summarize, in this paper we have shown that in the framework of 
an atom-laser 
approach via two coupled equations, one for the uncondensed phase
 and the other one for the  condensed phase,
addition of a diffusion term 
for the uncondensed atoms and 
 application of a proper adiabatic elimination procedure 
 leads to a CGL dynamical equation for an open BEC.
 In the case of negative scattering length, a suitable adjustment
 of the escape rate implies entering the unstable regime of the CGL dynamics.
 Furthermore, within the chosen ranges of the parameters 
 ($\varepsilon, c_1, c_2$), the 3-body recombination processes 
 have a negligible influence.

\acknowledgments

The Authors acknowledge E. Arimondo, 
 T. Hansch, L. Pitaevskii and the BEC group at LENS Firenze
 for fruitful discussions.
J.B. was partially supported by a EU Network grant (FMRXCT960010)
 "Nonlinear dynamics and statistical physics of spatially 
 extended systems" and by the Belgian Programme on 
Interuniversity Poles of Attraction (PAI 04-6) initiated by
the Belgian State Federal Office of Scientific, Technical and 
Cultural Affairs.
 This work was partially supported by INFM through the Advanced
 Research Project CAT.

\newpage

\centerline{{\bf CAPTIONS}}

{\bf Fig.1} Open BEC of Rb: Time evolution of $|\phi|^2$ 
using Eq.(\ref{eq20}) with 
$c_1=0.52$, $c_2=2.05$ and $\varepsilon=0.5$.
The time step 
is ($\Delta_t = 0.01$). The initial condition is a Gaussian.
The figures are coded using a grey scale (white corresponds
to the maximum value of $|\phi|^2$).\\

{\bf Fig.2} 
Cuts  of $|\phi|^2$ at $y=0$, same parameters as for Fig.1.
The dotted line corresponds to t=1, the dashed line is for
t=10 and the solid line (t=500) corresponds to the final 
stationary state.\\

{\bf Fig.3} Open BEC of Li$^*$
(escape parameters adjusted to be in the unstable regime):
Time evolution of $|\phi|^2$ using Eq.(\ref{eq20}) with
$c_1=0.4$, $c_2=-5.1$ and $\varepsilon=0.5$.\\

{\bf Fig.4} 
Cuts  of $|\phi|^2$ at $y=0$, same parameters as for Fig.3.
The dotted line corresponds to t=1, the dashed line is for
t=10 and the solid line (t=500) corresponds to a chaotic
state.\\

{\bf Fig.5} 
Spatial power spectrum of $|\phi(x,y=0,t)|^2$
for Li$^*$. The lower curve (dotted line)
corresponds to t=1. The upper curve (solid line) is calculated by averaging
the spatial power spectrum of  $|\phi(x,y=0,t)|^2$ between 
t=500 and t=1000 (by steps of $\delta t = 10$).\\

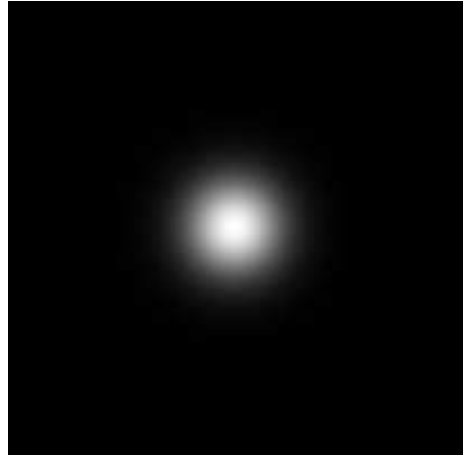
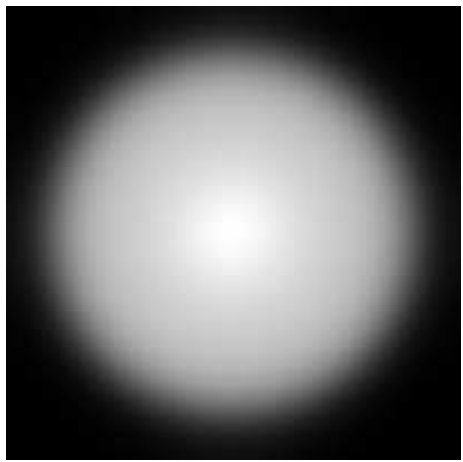
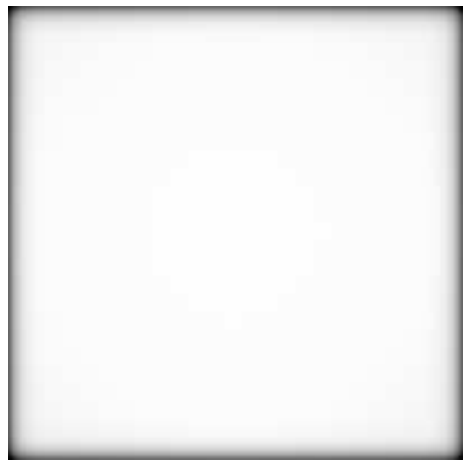
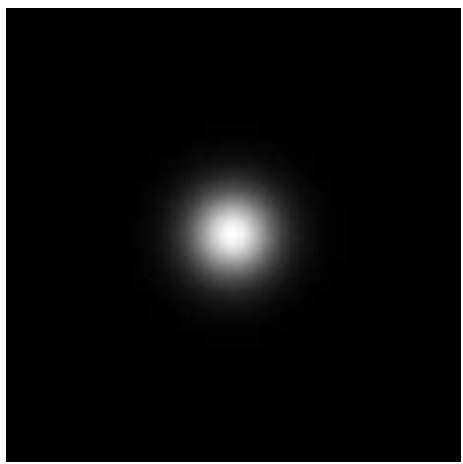
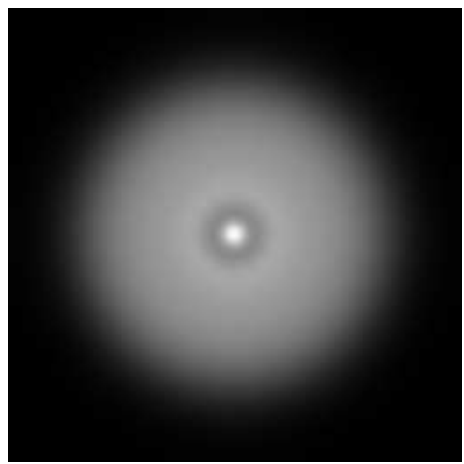
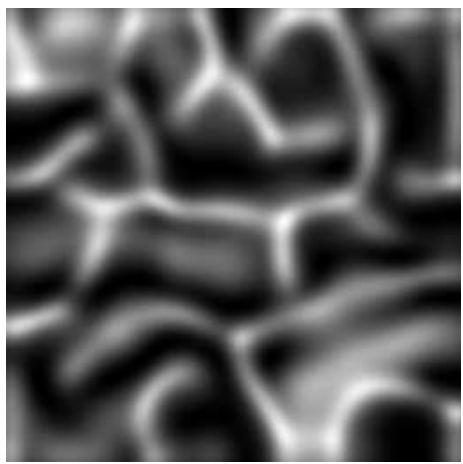
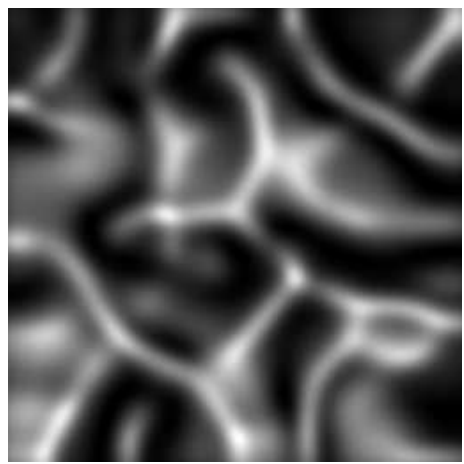
\end{document}